\documentclass[%
 reprint,
 amsmath,amssymb,
 aps,
 onecolumn,
]{revtex4-1}

\usepackage{graphicx}
\usepackage{dcolumn}
\usepackage{bm}
\usepackage{braket}

\usepackage{amsmath}	
\begin{document}

\preprint{APS/123-QED}

\title{Isoscalar dipole transition as a probe for asymmetric clustering}%

\author{Y. Chiba}
\author{M. Kimura}
\email{masaaki@nucl.sci.hokudai.ac.jp}
\affiliation{Department of Physics, Hokkaido University, Sapporo 060-0810, Japan}
\author{Y. Taniguchi}
\affiliation{Nihon Institute of Medical Science, Saitama 350-0435, Japan}

\date{\today}

\begin{abstract}
\begin{description}
\item[Background] The sharp $1^-$ resonances with enhanced isoscalar dipole transition strengths
	   are observed in many light nuclei at relatively small excitation energies, but their
	   nature was unclear.
\item[Purpose] We show those resonances can be attributed to the cluster states with asymmetric
	   configurations such as $\alpha$+$^{16}{\rm O}$. We explain why asymmetric cluster
	   states are strongly excited by the isoscalar dipole transition. We also provide 
	   a theoretical prediction of the isoscalar dipole transitions in $^{20}{\rm Ne}$ and
	   $^{44}{\rm Ti}$. 
\item[Method] The transition matrix is analytically derived to clarify the excitation
	   mechanism. The nuclear model  calculations by Brink-Bloch wave function and
	   antisymmetrized molecular dynamics are also performed to provide a theoretical
	   prediction for $^{20}{\rm Ne}$ and $^{44}{\rm Ti}$.   
\item[Results] It is shown that the transition matrix is as large as the Weisskopf estimate even
	   though the ground state is an ideal shell model state. Furthermore, it
	   is considerably amplified if the ground state has cluster correlation. The nuclear
	   model calculations  predict large transition matrix to the $\alpha$+$^{16}{\rm O}$ and
	   $\alpha$+$^{40}{\rm Ca}$ cluster states comparable with or larger than  the Weisskopf
	   estimate.   
 \item[Conclusion] We conclude that the asymmetric cluster states are strongly excited by the
	    isoscalar dipole transition. Combined with the isoscalar monopole transition that
	    populates the $0^+$ cluster states, the isoscalar transitions are promising probe for
	    asymmetric clusters. 
\end{description}
\end{abstract}

\pacs{Valid PACS appear here}
\maketitle


\section{introduction}
The observed electric monopole ($E0$) and isoscalar (IS) monopole strength distributions of light 
nuclei \cite{yb97,yb98-1,yb99,yb01,yb02,yb03-2,yb07,wk07,kw08,yb09-1,yb09-2,it11,
yb13,Gup15} show that considerable amount of the strength fractions appears at
relatively small excitation energy as sharp resonances. It was known that many of those resonances
are associated with the $\alpha$ cluster states such as the Hoyle state of  $^{12}{\rm C}$
\cite{Ueg77,Ueg79,Kam81,Des87,Kan98,Toh01,Fun03,Nef04,Fun06,Nef07}, 
the $\alpha$+$^{12}{\rm C}$ cluster states in $^{16}{\rm O}$ \cite{Suz89,Yam12},
$\alpha$+$^{16}{\rm O}$ cluster states in $^{20}{\rm Ne}$ \cite{Suz87},
 and $\alpha$+$\alpha$+$t$ cluster state in
$^{11}{\rm B}$  \cite{Kan07,Kaw07,Yam10}. Therefore, IS monopole transition has been utilized as a
probe to search for the cluster states in light nuclei. 
Later, Yamada, {\it et al.} \cite{Yam08} clearly explained the enhancement mechanism of the
monopole transition from the ground state to the cluster states. They showed, by exploiting the
Bayman-Bohr theorem \cite{Bay58}, the degrees-of-freedom of cluster excitation are embedded in the
ground state, and the monopole operator activates them to excite the cluster states strongly. This
finding boosted the studies of  the cluster states using the  monopole transition as a probe. In
these days,  various cluster states in stable and unstable nuclei 
\cite{Ich11,Ito11,it13,kw13,Kan14,Yan14,Yan15,Chi15} are discussed on
the basis of their enhanced monopole strengths.

Among various cluster states, the cluster systems with asymmetric configuration must have the
$1^-$ state that constitute the parity doublet together with the $0^+$ cluster state.
For example, the $1^-$ state of $^{16}{\rm O}$ at 9.6 MeV and that of $^{20}{\rm Ne}$ at 5.8 MeV
are  the evidence of the asymmetric clustering with $\alpha$+$^{12}{\rm C}$ and
$\alpha$+$^{16}{\rm O}$ configurations \cite{Hor68}.
Therefore,  identifying the  $1^-$ cluster state is a key to prove the asymmetric clustering.  In
the case of the $N\neq Z$ nuclei, the enhanced electric dipole ($E1$) transition was suggested as
a probe for such $1^-$ states \cite{Iac85}. The $1^-$ cluster states with $\alpha$+$^{14}{\rm C}$
configuration in $^{18}{\rm O}$ were experimentally searched by using the $E1$ strength as a
probe \cite{Gai82,Gai83-1}, in addition to the ordinary experimental methods such as
$\alpha$ scattering and breakup \cite{Cur02,Gol05,Ash06,Yil06}. The $\alpha$
clustering in the actinides \cite{Dal83,Gai83-2,Gai88} were also investigated in the same
literature. Recently, the $\alpha$ clustering in $^{212}{\rm Po}$ \cite{Ast10,Suz10,Del12} and the
Rare-Earth nuclei \cite{Spi15} were also discussed based on their enhanced low-lying $E1$
strengths.  

In the case of the $N=Z$ nuclei for which we expect a rich variety of clustering, the $E1$
transition is not available and we need other probes. As an alternative, one may consider the IS
dipole transition, because it populates $1^-$ states and has similar operator form to the IS
monopole  transition. Furthermore, its strength distributions measured for light nuclei 
\cite{yb01,yb02,yb03-2,yb09-1,yb09-2,yb13,it13,kw13}
show the existence of the sharp resonances with enhanced strengths at relatively small excitation
energies well below the giant resonance. In a recent experimental study \cite{it13}, the observed
low-lying resonances in $^{32}{\rm S}$ are conjectured to be the $\alpha$+$^{28}{\rm Si}$ cluster
states, because of their enhanced IS dipole transition strength from the ground state. Very
recently, Kanada-En'yo also discussed the enhancement of the IS dipole transition strength of the
$\alpha$ cluster states in $^{12}{\rm C}$ based on the theoretical calculation
\cite{Kan15}. However, the excitation mechanism of the IS dipole transition and the relationship
to the cluster states are still unclear, and must be clarified to promote the theoretical and 
experimental studies.

For this purpose, by using $\alpha$+$^{16}{\rm O}$ and $\alpha$+$^{40}{\rm Ca}$ cluster states in
$^{20}{\rm Ne}$ and $^{44}{\rm Ti}$ as examples, we investigate the excitation mechanism of the IS
dipole transition from the ground state to the asymmetric cluster states. We first discuss an
analytic expression of the IS transition matrix from the  shell model ground state to the cluster
states. It is found that the transition matrix is enlarged for asymmetric cluster systems and
becomes as large as the Weisskopf estimate, even if the ground state is an ideal shell model
state. Furthermore, a simple numerical estimate using Brink-Bloch wave function \cite{Bri66} shows 
that the matrix is considerably amplified if the ground state has cluster correlation.  

To provide more realistic prediction of the IS dipole transition strength in $^{20}{\rm Ne}$ and
$^{44}{\rm Ti}$, we also performed microscopic nuclear model calculations by using generator
coordinate method with Brink-Bloch wave function (Brink-Bloch GCM) and antisymmetrized molecular
dynamics (AMD) \cite{amd1,amd2}. AMD is able to describe the distortion of the clusters, and
reasonably explains the observed excitation spectra of the ground band and cluster bands for both
nuclei. The AMD result shows that only the $1^-$ states having $\alpha$+$^{16}{\rm O}$ or
$\alpha$+$^{40}{\rm Ca}$ cluster structure have large transition matrix comparable or larger than
the Weisskopf estimate and other non-cluster $1^-$ states are insensitive. From those results, we
conclude that the IS dipole transition  can  strongly excite the $1^-$ cluster states, and is a
promising probe for asymmetric clustering, when combined with the IS monopole transition that
excites $0^+$ cluster states.

This paper is organized as follows. In Sec. \ref{sec:estimate}, we derive an analytic
expression for the IS dipole transition matrix. We also perform simple numerical estimation of the
transition matrix using Brink-Bloch wave function. The microscopic models, Brink-Bloch GCM and AMD,
are introduced in Sec. \ref{sec:model}, and the results obtained by those models are discussed in
Sec. \ref{sec:mic}. The final section summarizes this study. 
\section{estimates of isoscalar dipole transition matrix}\label{sec:estimate}
In this section, using the shell model and cluster model wave functions, we estimate the magnitude of the
IS dipole transition matrix between the ground and excited $1^-$ states of $^{20}{\rm Ne}$
and $^{44}{\rm Ti}$ having asymmetric cluster structure with $\alpha$+${}^{16}{\rm O}$ and
$\alpha$+${}^{40}{\rm Ca}$ configurations. 

By assuming that the ground state is described by a shell model wave function, we first derive an
analytical expression for the transition matrix and show that it is comparable with the 
Weisskopf estimate. We also show that the transition matrix is considerably amplified
when the ground state has cluster correlation.

\subsection{Analytical estimate of the transition matrix}
\subsubsection{Wave functions of the ground state, \\nodal and angular excited cluster states}
The ground states of $^{20}$Ne and $^{44}$Ti are dominated by the $(0d1s)^4$ and $(0f1p)^4$
configurations on top of the closed shell cores $^{16}$O and $^{40}$Ca. The shell model
calculations \cite{Har69,Tom78,Var98,Mcg71,Mcg73,Pov81} showed that the ground state of $^{20}$Ne
is dominated by 
the $SU(3)$ irreducible representation of $(\lambda,\mu)$ = (8,0), and $^{44}$Ti is by the (12,0)
representation in the Elliott's $SU(3)$ model. An important fact here is that
these shell model wave functions are equivalently expressed by the $\alpha$+$^{16}$O and  
$\alpha$+$^{40}$Ca cluster model wave functions 
owing to the Bayman-Bohr theorem \cite{Bay58}.   
\begin{align}
 &\Phi({\rm gs}) = \frac{c_0}{\sqrt{\mu_{N_0}}}
 \mathcal A'\Set{\mathcal R_{N_000}(\bm r)\phi_{1}\phi_{2}},
 \label{eq:gswf}\\
 &c_0 =  \sqrt{C_1!C_2!/A!} ,\nonumber\\
 &\mu_N=\braket{\mathcal R_{Nlm}(\bm r)\phi_{1}\phi_{2}|
 \mathcal A'\Set{\mathcal R_{Nlm}(\bm r)\phi_{1}\phi_{2}}}\label{eq:rgmnorm}.
\end{align}
Here, the internal wave functions of $\alpha$ cluster (with mass $C_1$) and $^{16}$O or $^{40}$Ca
cluster (with mass $C_2$) denoted by $\phi_{1}, \phi_{2}$ are the harmonic oscillator wave
functions with the oscillator parameter $\nu=m\omega/(2\hbar)$. The wave function of the intercluster
motion is also the harmonic oscillator wave function  $\mathcal R_{Nlm}(\bm
r)=R_{Nl}(r)Y_{lm}(\hat r)$ but its oscillator parameter is scaled by the reduced mass
$\nu'=({C_1C_2}/{A})\nu$. The principal quantum number of the intercluster motion is equal to the
lowest Pauli allowed values, $N_0=8$ for $^{20}$Ne and $N_0=12$ for $^{44}$Ti (the nodal quantum
numbers $n_0 = (N_0 - l)/2$ are 4 and 6).    

As emphasized in Ref. \cite{Yam08}, this equivalence of the shell model and cluster model wave
functions implies that the degrees-of-freedom of cluster excitation are embedded even in
an ideal shell model ground state.
For example, the nodal excitation of the intercluster motion yields the excited
$0^+$ state,  
\begin{align}
 &\Phi(0^+_{\rm ex}) = \sum_{N=N_0+2}^{\infty} e_n \frac{c_0}{\sqrt{\mu_{N}}}
 \mathcal A'\Set{\mathcal R_{N00}(\bm r)\phi_{1}\phi_{2}},\label{eq:ex0wf}
\end{align}
where the nodal quantum number of intercluster motion is increased relative to the ground state,
and hence, the principal quantum number $N$ must be equal to or larger than $N_0+2$. Thus, the
states with larger values of $N$ are coherently superposed with coefficients $e_n$. The $0^+_4$
state of $^{20}$Ne around 8.7 MeV \cite{Til98} and $0^+$ states of $^{44}$Ti observed around 11
MeV \cite{Yam90,Yam96} are attributed to this class of nodal excited cluster state. In
Ref. \cite{Yam08}, taking $^{12}$C and $^{16}$O as examples, it was shown that the IS monopole
transition matrix from the ground state to the nodal excited cluster states is large.

Besides the nodal excitation,  the angular excitation of the intercluster motion
also takes place. For example, the angular excitation with $\Delta l=1$ (combined with
the nodal excitation) yields the $1^-$ state,  
\begin{align}
 &\Phi(1^-) = \sum_{N=N_0+1}^{\infty}  f_n \frac{c_0}{\sqrt{\mu_{N}}}
 \mathcal A'\Set{\mathcal R_{N10}(r)\phi_{1}\phi_{2}},\label{eq:ex1wf}
\end{align}
where the principal quantum number $N$ must be equal to or larger than $N_0+1$.
The $1^-_1$ state of $^{20}$Ne at 5.8 MeV \cite{Til98} and $1^-$ states of $^{44}$Ti observed at
6.2 MeV and around 12 MeV \cite{Yam90,Yam96,Yam93} are attributed to this class of angular excited
cluster state. Since the angular excitation with odd number angular momenta (negative-parity
states) is allowed only in the asymmetric cluster systems ($C_1\neq C_2$), the $1^-$ state has
been regarded as the evidence of the asymmetric clustering \cite{Hor68}.  
\subsubsection{Analytical expression of the transition matrix}
Using the wave functions described by Eqs. (\ref{eq:gswf}) and (\ref{eq:ex1wf}), we derive
an analytic expression for the IS dipole transition between the ground and 
the angular excited $1^-$ cluster states. The IS
dipole operator $\mathcal M_\mu^{ IS1}$,  reduced matrix element $M^{IS1}$ and transition
probability $B(IS1)$ are   
\begin{align}
 &\mathcal M_\mu^{ IS1}=\sum_{i=1}^{A}(\bm r_i - \bm r_{\rm cm})^2
 \mathcal Y_{1\mu}(\bm r_i-\bm r_{\rm cm}), \\
 &M^{IS1}=\braket{1^-||\mathcal M^{IS1}||0^+_1}
 = \sqrt{3}\braket{1^-,J_z|\mathcal M^{IS1}_{J_z}|0^+_1}, \label{eq:redmat}\\
 &B(IS1;0^+_1 \rightarrow 1^-) = |M^{IS1}|^2,
\end{align}
where $\bm r_i$ denotes the $i$th nucleon coordinate, while $\bm r_{\rm cm}$ denotes the
center-of-mass of the system. The solid spherical harmonics are
defined as $ \mathcal Y_{\lambda\mu} (\bm r)\equiv r^\lambda Y_{\lambda\mu}(\hat r)$.

Applying the  wave functions Eq. (\ref{eq:gswf}) and (\ref{eq:ex1wf}) to Eq. (\ref{eq:redmat}), the
reduced matrix element is given as,
\begin{align}
 &M^{IS1}=\sqrt{3}\braket{\Phi(1^-)|\mathcal M^{IS1}_0|\Phi({\rm gs})}=
\sum_{N=N_0+1}\frac{\sqrt{3}f_N}{\sqrt{\mu_{N_0}\mu_N}}\nonumber\\
 &\times\braket{\mathcal M_0^{IS1}\mathcal R_{N10}(\bm r)\phi_{1}\phi_{2}|\mathcal
 A'\Set{\mathcal R_{N_000}(\bm r)\phi_{1}\phi_{2}}}. \label{eq:redmat2}
\end{align}

To evaluate the last matrix element, we rewrite $\mathcal M_\mu^{IS1}$ in
terms of the internal coordinates $\bm \xi_i$ of each cluster and the intercluster coordinate $\bm r$
which are defined as,
\begin{align}
 &\bm R_{C_1} \equiv \frac{1}{C_1}\sum_{i\in C_1}\bm r_i,\quad
 \bm R_{C_2} \equiv \frac{1}{C_2}\sum_{i\in C_2}\bm r_i,\\
 &\bm \xi_i \equiv\left\{
 \begin{array}{c}
 \bm r_i - \bm R_{C_1},\quad i\in C_1\\
 \bm r_i - \bm R_{C_2},\quad i\in C_2\\
 \end{array}
\right.\\
 &\bm r \equiv \bm R_{C_1} - \bm R_{C_2}, \
\end{align}
where the center-of-mass of clusters $\bm R_{C_1}$ and $\bm R_{C_2}$ are introduced.
With these coordinates, as explained in appendix \ref{app:appa}, $\mathcal M_\mu^{IS1}$ is
expressed as,

\begin{widetext}
\begin{align}
 \mathcal M^{IS1}_\mu =&  \sum_{i\in C_1}\xi_i^2\mathcal Y_{1\mu}(\bm \xi_i)
 +\sum_{i\in  C_2}\xi_i^2\mathcal Y_{1\mu}(\bm \xi_i)
-\sqrt{\frac{32\pi}{9}}\Set{
 \frac{C_2}{A}\biggl[\sum_{i\in C_1}\mathcal Y_{2}(\bm \xi_i)
 \otimes \mathcal Y_{1}(\bm r)\biggr]_{1\mu}
 -\frac{C_1}{A}\biggl[\sum_{i\in C_2}\mathcal Y_{2}(\bm \xi_i)
 \otimes \mathcal Y_{1}(\bm r)\biggr]_{1\mu}
 }\nonumber\\
 &+\frac{5}{3}\left(
 \frac{C_2}{A}\sum_{i\in C_1}\xi_i^2 
 -\frac{C_1}{A}\sum_{i\in C_2}\xi_i^2 \right)\mathcal Y_{1\mu}(\bm r) 
 -\frac{C_1C_2(C_1-C_2)}{A^2}r^2\mathcal Y_{1\mu}(\bm r). \label{eq:isd_decompose}
\end{align}
\end{widetext}
This expression makes it clear that $\mathcal M^{\rm IS}_{\mu}$ will activate the
degrees-of-freedom of cluster excitation embedded in the ground state. It will generate angular
excited cluster states with $J^\pi=1^-$, because if operated to the ground state wave function
given in Eq. (\ref{eq:gswf}),  the terms depending on $\mathcal Y_{1\mu}(\bm r)$ and $r^2\mathcal 
Y_{1\mu}(\bm r)$ will induce the  nodal and angular excitation of the intercluster motion. 

By substituting Eq. (\ref{eq:isd_decompose}) into Eq. (\ref{eq:redmat2}), one finds that the first 
line of Eq. (\ref{eq:isd_decompose}) identically vanishes because it involves the internal
excitation of clusters. Hence, only the second line has non-vanishing matrix element as given
below  (see appendix \ref{app:appb} for the derivation).

\begin{widetext}
\begin{align}
 M^{IS1} =&
 \sqrt{\frac{3}{4\pi}}\frac{C_1C_2}{A}\Biggl[f_{N_0+1}\sqrt{\frac{\mu_{N_0}}{\mu_{N_0+1}}}
 \biggl\{
 \frac{5}{3}\left(\braket{r^2}_{C_1} - \braket{r^2}_{C_2}\right)
\braket{R_{N_00}|r|R_{N_0+11}}
 -\frac{C_1-C_2}{A}\braket{R_{N_00}|r^3|R_{N_0+11}}\biggr\} \nonumber\\
 &\qquad\qquad\qquad -\frac{C_1-C_2}{A}
 f_{N_0+3}\sqrt{\frac{\mu_{N_0}}{\mu_{N_0+3}}}\braket{R_{N_00}|r^3|R_{N_0+31}}\Biggr],
 \label{eq:redmat5}
\end{align}
\end{widetext}
where $\braket{r^2}_{C_1}$ and $\braket{r^2}_{C_2}$ are the square of the root-mean-square radius
of the clusters, and  
the matrix elements of harmonic oscillator are given as,
\begin{align}
&\braket{R_{N_00}|r|R_{N_0+11}} = \sqrt{\frac{N_0+3}{4\nu'}},\nonumber\\
&\braket{R_{N_00}|r^3|R_{N_0+11}} = \frac{3N_0+5}{4\nu'}\sqrt{\frac{N_0+3}{4\nu'}},\nonumber\\
&\braket{R_{N_00}|r^3|R_{N_0+31}} = -\frac{\sqrt{(N_0+2)(N_0+5)}}{4\nu'}
 \sqrt{\frac{N_0+3}{4\nu'}}.  \label{eq:ho1}
\end{align}
From Eq. (\ref{eq:redmat5}), we find following  properties. (1) The transition matrix  is
proportional to $\braket{r_{C_1}^2}- \braket{r_{C_2}^2}$ or $(C_1-C_2)/A$, which means that it is
amplified for the asymmetric cluster states. Therefore, we expect IS dipole transition is a good
probe for asymmetric clustering.  (2) For the cluster states dominated by $1\hbar\omega$
configuration, the first line of Eq. (\ref{eq:redmat5}) dominantly contributes, while the second
line becomes major for the $3\hbar\omega$ excited cluster states.

\subsubsection{Estimation of the matrix element}
We are now able to estimate the magnitude of the IS dipole transition matrix. We adopted
the values listed in 
Table  \ref{tab:value}. Here, the oscillator parameters $\nu=0.16$ $\rm fm^{-2}$ for $^{20}$Ne
and $0.12$ $\rm fm^{-2}$ for $^{44}$Ti are so determined to minimize the ground state energies as
explained in Sec. \ref{sec:model}. The coefficients $f_{N_0+1}$ and $f_{N_0+3}$ are estimated
by the AMD calculation which is also explained in Sec. \ref{sec:model}. For other quantities,  
analytical calculation is possible or experimental value is available.
\begin{table}[h]
\caption{List of the quantities used to evaluate Eq. (\ref{eq:redmat5}). Radii of  $\alpha$,
 $^{16}$O and $^{40}$Ca clusters are calculated from the measured charge radii given in
 Ref. \cite{Ang11}  and  listed in the units of   ${\rm fm}^2$. The oscillator parameters $\nu$
 and $\nu'$ are in  units of ${\rm  fm}^{-2}$. Other  
 quantities are  dimensionless.}\label{tab:value} 
\begin{ruledtabular}
\begin{tabular}{ccccccc}
 &$N_0$& $\mu_{N_0}$
 \footnote{$\mu_N$ defined in Eq. (\ref{eq:rgmnorm}) is so-called eigenvalue of RGM norm kernel
 and analytically calculable. The  values listed in the table are taken from Ref. \cite{Hor77}.}
 & $\mu_{N_0+1}$&$\mu_{N_0+3}$& $\braket{r^2}_{C_1}$ &$\braket{r^2}_{C_2}$\\
\hline
 $^{20}{\rm Ne}$ & 8 & 0.229 & 0.344&0.620&
 $(1.46)^2$&$(2.57)^2$\\
 $^{44}{\rm Ti}$ & 12& 0.069 & 0.157&0.372&
 $(1.46)^2$&$(3.37)^2$\\
\hline
 & $\nu$ & $\nu'$ &  $f_{N_0+1}$ &$f_{N_0+3}$& &\\
\hline
 $^{20}{\rm Ne}$ & 0.16 & 0.51 &$\sqrt{0.39}$ &-$\sqrt{0.28}$  & &\\
 $^{44}{\rm Ti}$ & 0.12 & 0.44 &$\sqrt{0.23}$ &-$\sqrt{0.26}$  & &\\
\end{tabular}
\end{ruledtabular}
\end{table}
Assignment of those values to Eq. (\ref{eq:redmat5}) yields the estimation for $^{20}{\rm Ne}$, 
\begin{align}
 M^{IS1}(^{20}{\rm Ne}) = 3.08 f_{9} -7.36 f_{11} = 5.82\ \rm fm^3, \label{eq:wune}
\end{align}
and for $^{44}{\rm Ti}$,
\begin{align}
 M^{IS1}(^{44}{\rm Ti}) = 13.3 f_{13} - 16.2 f_{15} = 14.6\ \rm fm^3.  \label{eq:wuti}
\end{align}
It is noted that $f_{N_0+1}$ and $f_{N_0+3}$ usually have opposite sign for cluster states as
explained in appendix \ref{app:appc}, and hence, the first and second terms in
Eqs. (\ref{eq:wune}) and (\ref{eq:wuti}) are additively contribute to enlarge the matrix element.

These results are compared with the single-particle estimates. Assuming the constant radial wave
function as usual, Weisskopf estimate is given as 
\begin{align}
 M_{\rm WU}^{IS1} = \sqrt{\frac{3}{4\pi}}\frac{3}{6}(1.2A^{1/3})^3 \simeq 0.422 A\ \rm fm^3.
\end{align}
It is approximately 8.44 ${\rm fm}^3$ for $^{20}{\rm Ne}$ and 18.6 ${\rm fm}^3$ for
$^{44}{\rm Ti}$, which are slightly larger than but comparable with Eqs. (\ref{eq:wune}) and
(\ref{eq:wuti}).  

Thus, the angular excited cluster states have strong IS dipole transition from the ground
state comparable with the Weisskopf estimate, even if the ground state is not a cluster state but
an ideal shell model state. Since the single-particle transition is usually fragmented into many
states, only the asymmetric cluster states can have strong transition strengths. Furthermore, as
we will show below, the strength is further amplified if the ground state has cluster correlation.


\subsection{Amplification of the transition matrix\\ owing to the clustering of the ground state}\label{sec:estimate:amp}
Here we show that the magnitude of $M^{IS1}$ is considerably amplified compared to the
estimates made in the previous subsection, if the ground state has cluster correlation. To
demonstrate it, we employ  Brink-Bloch wave function \cite{Bri66} that is composed  of clusters
$C_1$ and $C_2$ placed at $-C_2/A\bm D$ and $C_1/A\bm D$ with the intercluster distance $D$,  
\begin{align}
 &\Phi_{\rm BB}(D) = n_0\mathcal A'\left\{
 \psi_{C_1}\Bigl(-\frac{C_2}{A}\bm D\Bigr) \psi_{C_2}\Bigl(\frac{C_1}{A}\bm D\Bigr)
 \right\},\label{eq:BB1}\\
 &\bm D = (0,0,D),\nonumber
\end{align}
where $\psi_{C_1}$ and $\psi_{C_2}$ denote the wave functions of clusters represented by the
harmonic oscillator wave functions that include their center-of-mass coordinates . The oscillator 
parameter are $\nu=0.16$ and $0.12$ $\rm fm^{-2}$ for $^{20}$Ne and $^{44}$Ti, respectively. 
Eq. (\ref{eq:BB1})  is projected to the eigenstate of parity and angular momentum, 
\begin{align}
 &\Phi^\pi_{\rm BB}(D) = \frac{1+\pi P_x}{2}\Phi_{\rm BB}(D),\quad
 \pi=\pm,\\
 &\Phi^{l\pi}_{\rm BB}(D) = \frac{2l+1}{8\pi^2}\int d\Omega D^{l*}_{M0}(\Omega)
 R(\Omega)\Phi^\pi_{\rm BB}(D). \label{eq:BB3}
\end{align}
Here $P_x$, $D^l_{MK}(\Omega)$ and $R(\Omega)$ denote parity operator, Wigner $D$ function and
rotation operator. It is known that Brink-Bloch wave function can be transformed into the form of 
Eqs. (\ref{eq:gswf}), (\ref{eq:ex0wf}) and (\ref{eq:ex1wf}) \cite{Hor77}, 
\begin{align}
 &\Phi_{\rm BB}^{l\pi}(D) = \phi_{\rm cm}(\bm r_{\rm cm})\cdot n_0 \mathcal A'
 \set{\chi_{\rm BB}(\bm r)\phi_{1}\phi_{2}},\label{eq:BB2}\\
 &\phi_{\rm cm} = \left(\frac{2A\nu}{\pi}\right)^{3/4}e^{-A\nu r_{\rm cm}^2},\nonumber\\
 &\chi_{\rm BB}(\bm r) 
 =\sum_{N}A_{Nl}  \frac{(\nu' D^2)^{N/2}}{\sqrt{N!}}e^{-\nu'D^2/2}
 \mathcal R_{Nl0}(\bm r), \label{eq:BB4}\\
 &A_{Nl} = (-)^{(N-l)/2}\sqrt{\frac{(2l+1)N!}{(N-l)!!(N+l+1)!!}}. \label{eq:BB5}
\end{align}
where $\phi_{\rm cm}(\bm r_{\rm cm})$ is the center-of-mass wave function, and the wave
function of the intercluster motion $\chi_{\rm BB}(\bm r)$ is expanded by the harmonic oscillator
wave functions. From this  expression, we can see that Brink-Bloch wave function becomes identical
to the Eq. (\ref{eq:gswf}) at the limit of $D\rightarrow 0$, and hence, equals to the shell model
wave function. Of course, as $D$ increases, the wave function exhibits stronger clustering.  

\begin{figure}[t]
  \includegraphics[width=0.7\hsize]{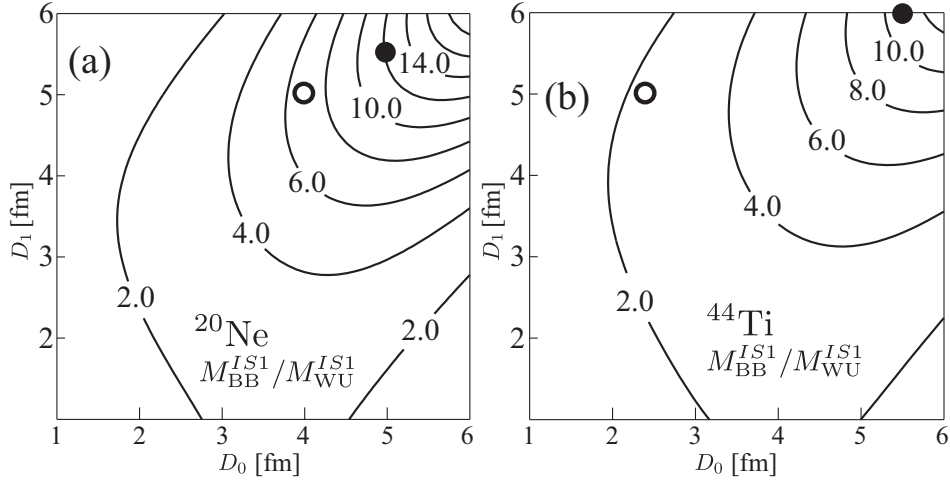}
  \caption{The ratio of the transition matrix to the Weisskopf estimate 
 $M_{\rm BB}^{IS1}/M_{\rm WU}^{IS1}$ as function of the intercluster distances 
 in the ground state ($D_{0}$) and in the $1^-$ state ($D_{1}$). The panel (a) is for 
 $\alpha$+$^{16}{\rm O}$ $({}^{20}{\rm Ne})$ system, while the panel (b) is for
 $\alpha$+$^{40}{\rm Ca}$ $({}^{44}{\rm Ti})$ system. The circles show the  approximate  positions
 of the  ground and the excited $1^-$ states obtained by the Brink-Bloch GCM (filled circles) and
 AMD (open circles) calculations given in the Sec. \ref{sec:mic}.}\label{fig:isd_brink}  
\end{figure}

Using the  Brink-Bloch wave functions for $\alpha$+${}^{16}{\rm O}$ $({}^{20}{\rm Ne})$ and
$\alpha$+${}^{40}{\rm Ca}$ $({}^{44}{\rm Ti})$ systems, we calculated the transition
matrix,  
\begin{align}
 &M^{IS1}_{\rm BB}(D_{0},D_{1}) = 
  \frac{\sqrt{3}
 \braket{\Phi_{\rm BB}^{1^-}(D_{1})|\mathcal M^{IS1}_0|\Phi_{\rm BB}^{0^+}(D_{0})}}
 {\sqrt{\braket{\Phi_{\rm BB}^{0^+}(D_{0})|\Phi_{\rm BB}^{0^+}(D_{0})}
 \braket{\Phi_{\rm BB}^{1^-}(D_{1})|\Phi_{\rm BB}^{1^-}(D_{1})}}}.
\end{align}
The result is shown in Fig. \ref{fig:isd_brink} where the ratio of the transition matrix to the
Weisskopf estimates of Eqs. (\ref{eq:wune}) and (\ref{eq:wuti}) are plotted as
functions of the intercluster distances $D_{0}$ in the ground state and
$D_{1}$ in the $1^-$ state. 
In both systems,  we see that even for the small values of $D_{0}$ and $D_{1}$,
$M_{\rm BB}^{IS1}$ is larger than the Weisskopf estimates. It is impressive that the 
matrix element is considerably amplified, as both of $D_{0}$ and $D_{1}$ increase. 

By more detailed calculations explained in the next section, the position of the ground and $1^-$
states are estimated approximately at the open circles in
Fig. \ref{fig:isd_brink}. Therefore, the transition strength is indeed considerably amplified and
regarded as a good probe for asymmetric clustering.

\section{Microscopic nuclear models}  \label{sec:model}
To provide realistic and reliable results for the IS dipole transition of $^{20}$Ne and
$^{44}$Ti, we performed two microscopic nuclear model calculations which we explain in this
section. The first is the Brink-Bloch GCM and the other is AMD. Brink-Bloch GCM can describe the  
intercluster motion properly. In addition to this, AMD can also describe the the
polarization and distortion of clusters. 

In both of theoretical models, the following microscopic $A$-body Hamiltonian is commonly used. 
\begin{align}
 {H} = \sum_{i=1}^A {t}(i) + \sum_{i<j}^A {v}_n(ij) + \sum_{i<j}^Z {v}_C(ij) - {t}_{\rm cm},
\end{align}
where the Gogny D1S interaction \cite{gogn91} is used as an effective nucleon-nucleon interaction
$v_n$. Coulomb interaction $v_C$ is approximated by a sum of seven
Gaussians. The center-of-mass kinetic energy ${t}_{\rm cm}$ is exactly removed. 

\subsection{Generator coordinate method\\ with Brink-Bloch wave function}
The Brink-Bloch GCM uses Eq. (\ref{eq:BB3}) as the basis function and employs the intercluster
distance $D$ as generator coordinate. The width parameter $\nu$ is so chosen to minimize the
ground state energy, that is found to be $\nu=0.16$ and $0.12$ $\rm fm^{-2}$ for $^{20}$Ne and
$^{44}$Ti, respectively. In the practical calculation, $D$ is discretized ranging from 1.0  to
12.0 $\rm fm$ with an interval of 0.5 $\rm fm$, that generates 23 basis functions
$\Phi^{l\pi}_{\rm BB}(D_i), \ i=1,...,23$.

To describe the ground and $\alpha$+$^{16}{\rm O}$ cluster states, the basis functions are
superposed, 
\begin{align}
 &\Psi^{l\pi}_{Mp} = \sum_{i}g_{ip}\Phi^{l\pi}_{BB}(D_i),
\end{align}
By solving the following Griffin-Hill-Wheeler equation \cite{Hill53,Hill57}, we obtain the
eigenenergy $E_p$ and the coefficients of the superposition $g_{ip}$.
\begin{align}
 &\sum_{i'}{H^{l\pi}_{ii'}g_{i'p}} = E^{l\pi}_p \sum_{i'}{N^{l\pi}_{ii'}g_{i'p}},\\
 &H^{l\pi}_{ii'} = \langle{\Phi^{l\pi}_{\rm BB}(D_i)|\hat{H}|\Phi^{l\pi}_{\rm BB}(D_{i'})}\rangle, \\
 &N^{l\pi}_{ii'} = \langle{\Phi^{l\pi}_{\rm BB}(D_i)|\Phi^{l\pi}_{\rm BB}(D_{i'})}\rangle.
\end{align}
Using thus-obtained wave functions for the ground and excited $1^-$ states, the reduced matrix
element given in Eq. (\ref{eq:redmat}) is directly calculated. 

\subsection{Antisymmetrized molecular dynamics}
In the AMD model \cite{amd1,amd2}, each nucleon is represented by a localized Gaussian wave
packet,  
\begin{align}
 \varphi_i({\bm r}) &= \exp\biggl\{-\sum_{\sigma=x,y,z}\nu_\sigma
  \Bigl(r_\sigma - 
  \frac{Z_{i\sigma}}{\sqrt{\nu_\sigma}}\Bigr)^2\biggr\}\chi_i\xi_i, \label{eq:singlewf}\\
 \chi_i &= a_i\chi_\uparrow + b_i\chi_\downarrow,\quad
 \xi_i = {\rm proton} \ {\rm or} \ {\rm neutron},\nonumber
\end{align}
where $\chi_i$ and $\xi_i$ represent spin and isospin wave functions. The intrinsic wave function
is a Slater determinant of nucleon wave packets,   
\begin{align}
  \Phi_{\rm int} ={\mathcal A} \{\varphi_1\varphi_2...\varphi_A \}.  \label{eq:wf_amd}
\end{align}
The parameters of the intrinsic wave function, ${\bm Z}_i$, $a_i$, $b_i$ and $\nu_\sigma$, are
determined by the energy minimization explained below.

Before the energy minimization, the intrinsic wave function is projected to the eigenstate of the
parity, 
\begin{align}
 \Phi^\pi &= \frac{1+\pi P_x}{2}\Phi_{\rm int},\quad \pi = \pm,
  \label{EQ_INTRINSIC_WF}  
\end{align}
Then, the above-mentioned parameters  are determined to minimize the expectation value of the
Hamiltonian $\widetilde{E}$ that is defined as  
\begin{align}
 \widetilde{E}&=\frac{\langle \Phi^\pi|\hat H|\Phi^\pi\rangle}{\langle
  \Phi^\pi|\Phi^\pi\rangle} + V_c,\\
 V_c&=v_\beta(\langle\beta\rangle-\beta_0)^2
  +v_\gamma(\langle\gamma\rangle-\gamma_0)^2.
\end{align}
Here the potential $V_c$ is added to impose the constraint on the quadrupole deformation of
intrinsic wave function that is parameterized by $\braket{\beta}$ and $\braket{\gamma}$  as
defined  in Ref. \cite{Kim12}. The magnitudes of  $v_\beta$ and $v_\gamma$ are chosen
large enough so that $\braket{\beta},\braket{\gamma}$ are, after the energy minimization, 
equal to $\beta_0,\gamma_0$.  By the energy minimization, we obtain the optimized wave
function $\Phi^{\pi}_{\rm int}(\beta_0,\gamma_0)$ for discretized sets of $(\beta_0, \gamma_0)$ on
the triangular lattice in $\beta$-$\gamma$ plane. The lattice size  is 0.05 and the calculation
is performed up to $\beta=0.9$.

After the energy minimization, we project out an eigenstate of angular momentum and perform
the GCM calculation by using quadrupole deformation parameters $\beta_0,\gamma_0$ as the generator
coordinates. We also included the Brink-Bloch wave functions $\Phi^{J\pi}_{\rm BB}(D_i)$ as the 
basis  functions of GCM. For simplicity, we denote by $\Phi_i$ this set of basis
functions. Because the AMD wave function is not necessarily axially symmetric,  non-zero values 
of $K$ quantum number and their mixing must be taken into account. Hence the equation for 
the angular momentum projection and Griffin-Hill-Wheeler equation are 
\begin{eqnarray}
 \Phi^{J\pi}_{MKi} = \frac{2J+1}{8\pi^2}\int d\Omega
  D^{J*}_{MK}(\Omega)\hat{R}(\Omega)\Phi^{\pi}_i,
\end{eqnarray} 
and
\begin{align}
 &\Psi^{J\pi}_{Mp} = \sum_{Ki}g_{Kip}\Phi^{J\pi}_{MKi},\label{eq:gcmwf}\\
 &\sum_{i'K'}{H^{J\pi}_{KiK'i'}g_{K'i'p}} = E^{J\pi}_p
 \sum_{i'K'}{N^{J\pi}_{KiK'i'}g_{K'i'p}},\\ 
 &H^{J\pi}_{KiK'i'} = \langle{\Phi^{J\pi}_{MKi}|\hat{H}|\Phi^{J\pi}_{MK'i'}}\rangle, \\
 &N^{J\pi}_{KiK'i'} = \langle{\Phi^{J\pi}_{MKi}|\Phi^{J\pi}_{MK'i'}}\rangle.
\end{align}
Using the wave function given in Eq. (\ref{eq:gcmwf}) the transition matrix element is calculated.

For a better understanding of the results presented in the next section, it is helpful to note
the differences between the Brink-Bloch GCM and AMD.  First, because nucleons are treated as
independent wave packets, AMD is able to describe various non-cluster states as well as the
cluster states, while Brink-Bloch GCM is not. Secondly, from the same reason, AMD is capable to
describe the polarization and distortion of clusters. Finally, since the Brink-Bloch wave
functions are also employed as the basis function, the AMD includes the Brink-Bloch GCM as a part
of its model space.  In short, in the AMD, the distortion of clusters and the coupling between the
cluster states and non-cluster states are taken into account. 

\subsection{Projection of AMD wave function to Brink-Bloch wave function}
As explained above, AMD wave function is admixture of the cluster and non-cluster wave functions. 
To identify the cluster state from AMD results, it is convenient to introduce an approximate
projector to Brink-Bloch wave function,
\begin{align}
& P_{\rm BB} = \sum_{ij} \ket{\Phi^{J\pi}_{\rm BB}(D_i)}(B^{-1})_{ij}\bra{\Phi^{J\pi}_{\rm
 BB}(D_j)}, 
\end{align}
where $B^{-1}$ is the inverse of overlap matrix $B$ defined 
$B_{ij} = \braket{\Phi^{J\pi}_{\rm BB}(D_i)|\Phi^{J\pi}_{\rm BB}(D_j)}$. With this projector, AMD
wave function Eq. (\ref{eq:gcmwf}) is projected to Brink-Bloch wave function, 
\begin{align}
 &P_{\rm BB}\Psi_{Mp}^{J\pi} = \sum_i G_i\Phi^{J\pi}_{\rm BB}(D_i),\label{eq:prj1}\\
 &G_i = \sum_{j}(B^{-1})_{ij}\braket{\Phi_{\rm BB}^{J\pi}(D_i)|\Psi_{Mp}^{J\pi}}
\end{align}
By substituting Eq. (\ref{eq:BB2}), (\ref{eq:BB4}) and (\ref{eq:BB5}) into r.h.s of
Eq. (\ref{eq:prj1}) and by comparing it with Eq. (\ref{eq:ex1wf}), we  calculated the coefficient
of superposition $f_{N_0+1}$ and $f_{N_0+3}$ given in Tab. \ref{tab:value}. 
The projector is also used to evaluate the amount of the cluster component in the AMD wave
function, that is defined as
\begin{align}
 S = \braket{\Psi^{J\pi}_{Mp}|P_{\rm BB}|\Psi^{J\pi}_{Mp}}.\label{eq:overlap}
\end{align}
When this value is sufficiently large, the excited state may be regarded as cluster state. 

We also explain how we estimated the intercluster distances $D_{0}$ and $D_{1}$ which are
shown by circles in Fig. \ref{fig:isd_brink}. We calculate the overlap between the GCM wave
functions $\Psi_{Mp}^{J\pi}$ for the ground and $1^-$ states and Brink-Bloch wave function
$\Phi_{\rm BB}^{J\pi}(D_i)$,  
\begin{align}
\frac{|\braket{\Psi_{Mp}^{J\pi}|\Phi_{\rm BB}^{J\pi}(D_i)}|^2}{
\braket{\Phi_{\rm BB}^{J\pi}(D_i)|\Phi_{\rm BB}^{J\pi}(D_i)}},\nonumber
\end{align}
and regard the distance $D_i$ at which the overlap has its maximum as $D_{0}$ or $D_{1}$.

\section{microscopic model calculations for isoscalar dipole transition}\label{sec:mic}
In this section, we discuss the IS dipole transitions in $^{20}{\rm Ne}$ and $^{44}{\rm Ti}$ 
studied by Brink-Bloch GCM and AMD. In Refs. \cite{Kim04,Tan04,Kim06}, the cluster and non-cluster
states of $^{20}{\rm Ne}$ and $^{44}{\rm Ti}$ have already been discussed based on AMD
calculation, and the reader is directed to those references for more detail. Here we focus
on the $\alpha$+$^{16}{\rm O}$  and  $\alpha$+$^{40}{\rm Ca}$  cluster states and discuss the
IS dipole transitions from the ground state to those cluster states.

\subsection{$\bm \alpha+\bf {}^{16}{O}$ cluster states in $^{\bf 20}{\bf Ne}$}
The $\alpha$+$^{16}{\rm O}$ cluster states in $^{20}{\rm Ne}$ have been studied in detail by many
authors \cite{Mat75,Hee76,Kru92,Buc75,Hee76,Fuj79,Kru92,Duf94,Tak96,Kim04,Tan04,Bo13,Zhou15}
 and well established.
\begin{figure}[h]
  \includegraphics[width=0.5\hsize]{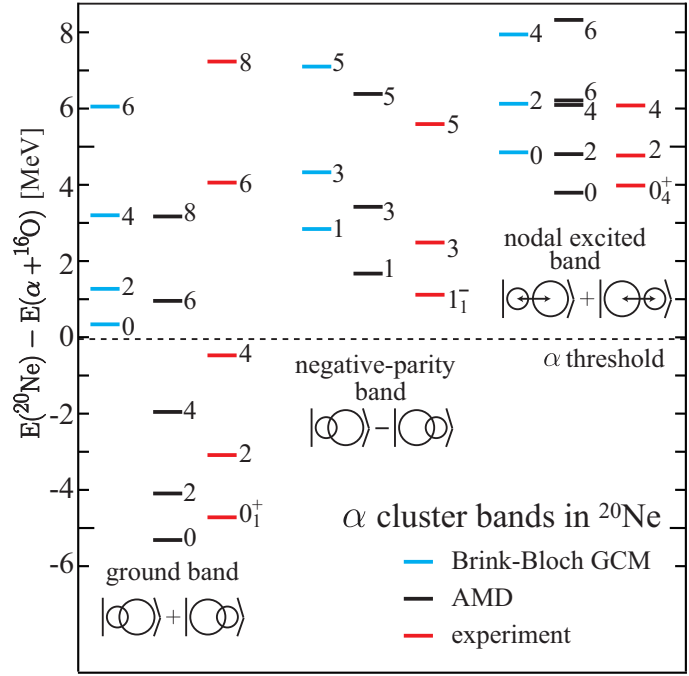}
  \caption{(color online) The observed and calculated $\alpha$+$^{16}{\rm O}$ cluster states in 
 $^{20}{\rm Ne}$ classified into three rotational bands. Energy is measured from the
 $\alpha$ threshold located at 4.7 MeV above the ground state. In the AMD result,
 the $6^+$ member state of nodal excited band is fragmented into two  states due to the coupling 
 with non-cluster configurations.}\label{fig:Ne20level}   
\end{figure}

The observed $\alpha$+$^{16}{\rm O}$ cluster bands are summarized in Fig. \ref{fig:Ne20level}
together with the results of Brink-Bloch GCM and AMD. The $0^+_4$ state observed at 8.7 MeV (4 MeV
above the $\alpha$ threshold) has  large $\alpha$ decay width  comparable with the
Wigner limit and is known as the nodal excited cluster state described by the wave function of 
Eq. (\ref{eq:ex0wf}). A rotational band is built on this state, which hereafter we call ``nodal
excited band''.  
The $1^-_1$ state at 5.8 MeV (1.1 MeV above the $\alpha$
threshold) also has large $\alpha$ decay width and is known as the angular excited 
cluster state described by the wave function of Eq. (\ref{eq:ex1wf}). This $1^-_1$ state is of
particular importance because it is regarded as the evidence for the asymmetric clustering with
$\alpha$+$^{16}{\rm O}$ configuration. On this state, the negative-parity band is built. 

It is well known that the ground band is the positive-parity partner of the
negative-parity band, and those two bands constitute the parity doublet \cite{Hor68}. This means
that the ground state has non-negligible cluster  correlation. 
Therefore, on the basis of the discussion made in in Sec. \ref{sec:estimate:amp}, we
expect that the IS dipole transition to the $1^-_1$ state is considerably amplified. 

\begin{table}[h]
\caption{The estimated intercluster distance of the ground and the $\alpha$+$^{16}{\rm O}$ cluster
states in  units of fm, and the IS dipole and monopole transition matrix from the ground
state to the $1^-_1$ and $0^+_4$   states in units of $\rm fm^3$ and $\rm fm^2$. Numbers in parenthesis
are ratio to the Weisskopf  estimates. }\label{tab:Ne20ism} 
\begin{ruledtabular}
\begin{tabular}{cccccc}
 &$D_0$& $D_1$ & $D(0^+_4)$&$M^{IS1}$ & $M^{IS0}$  \\
\hline
 BB GCM &  5.0   & 5.5 & 6.5 & 90.2 (10.7) & 46.4  (7.3) \\
 AMD    & 4.0    & 5.0 & 6.0 & 38.0 (4.5) & 16.0  (2.5) \\
\end{tabular}
\end{ruledtabular}
\end{table}

Then, we examine theoretical results. In the case of $^{20}{\rm Ne}$, it was easy to identify the
$\alpha$+$^{16}{\rm O}$ cluster states from AMD results, because all of the states shown in
Fig. \ref{fig:Ne20level} have large values of $S$ defined in Eq. (\ref{eq:overlap}). For example, 
$S=0.69, 0.90$ and 0.81 for the ground, $1^-_1$ and $0^+_4$ states. 

It is interesting to note the difference between the  Brink-Bloch GCM and AMD results. The
Brink-Bloch GCM fails to reproduce the energy of the ground band, while AMD reasonably describes
it, that indicates the importance of the cluster distortion effect. Indeed the estimated
intercluster distance $D$ is reduced in AMD compared to Brink-Bloch GCM  as listed in
Tab. \ref{tab:Ne20ism}. On the other hand,  both 
theoretical models give reasonable description for negative-parity and nodal excited
bands. Therefore, we can regard that the distortion effect is less important and almost ideal
clustering is realized in the $0^+_4$ and $1^-_1$ states, for which  both theoretical models
yielded large intercluster distances. 

The calculated IS dipole transition matrix from the ground state to the $1^-$ states are 
listed in Tab. \ref{tab:Ne20ism}. It is evident that the transition is greatly enhanced compared
to Weisskopf estimates. In particular, Brink-Bloch GCM yielded huge values, that is due to
the too weak binding of the ground state leading to the overestimation of the radius and cluster
correlation of the ground state. If the cluster distortion effect is taken into account by AMD,
the strength is somewhat reduced but still much larger than the Weisskopf estimate. We also note
that the nodal excited state ($0^+_4$) has large monopole transition matrix as
expected. Therefore, the present results suggest that both of the positive- and negative-parity
cluster states can be strongly generated by the IS monopole and dipole transitions from the
ground state, and hence, those transitions will be good signature of the asymmetric clustering.

\subsection{$\bm \alpha+\bf {}^{40}{Ca}$ cluster states in $^{\bf 44}{\bf Ti}$}
The $\alpha$+$^{40}{\rm Ca}$ cluster states in $^{44}{\rm Ti}$ have also been studied by many
authors
\cite{Yam90,Yam93-2,Yam96,Yam93,Joh69,Str74,Fre75,Fre76,Fre83,Cha84,Mic86,Wad88,Ohk98,Kim06}, but 
the situation is more complicated than the case of  $^{20}{\rm Ne}$. The theoretical and
experimental studies are summarized in a review paper \cite{OhkR}, and we discuss based on the
assignment given therein. 
\begin{figure}[h]
 \includegraphics[width=0.5\hsize]{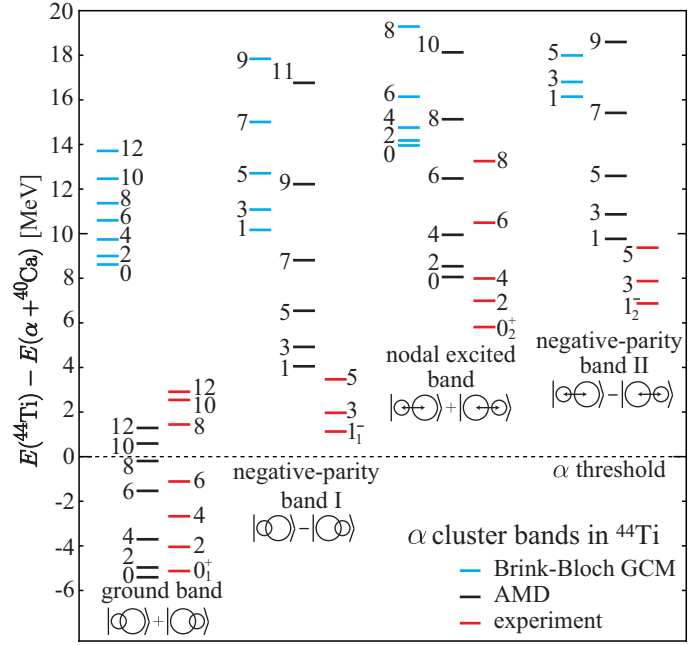}
 \caption{(color online) The observed and  calculated $\alpha$+$^{40}{\rm Ca}$ cluster states in
 $^{44}{\rm Ti}$  classified into four  rotational bands. Energy is measured from the  $\alpha$
 threshold located  at 5.1 MeV above the  ground state. In the experiment and AMD results, the
 member states of the   nodal excited band and negative-parity band II are fragmented into several
 levels, and the  weighted averages of those levels are shown in the figure.}\label{fig:Ti44level}    
\end{figure}

Figure \ref{fig:Ti44level} shows the observed candidates of $\alpha$+$^{40}{\rm Ca}$ cluster
states together with the present theoretical results. Based on the $\alpha$ transfer
experiment, four rotational bands including the ground band were classified as the
$\alpha$+$^{40}{\rm Ca}$ cluster bands. The $1^-$ state observed at 6.2 MeV (1.1 MeV above the
$\alpha$ threshold) is strongly populated by the $\alpha$ transfer reaction \cite{Yam93, Yam93-2}
and is the angular excited cluster state dominated by the 1$\hbar\omega$ excitation of the
intercluster motion.  Although it is not the yrast $1^-$ state, we call this state the $1^-_1$
state in the following. On this $1^-_1$ state, a rotational band which we call ``negative-parity
band I'' is built. 

A couple of candidates of the nodal excited $0^+$  state are reported around 11.0 MeV (5.9
MeV above the $\alpha$ threshold) by the $\alpha$ elastic scattering \cite{Fre83} and
the $\alpha$ transfer reaction \cite{Yam90, Yam96, Yam93-2}. Those data suggest that the nodal
excited   cluster state may be fragmented into several states due to the coupling with other
non-cluster configurations. In Fig. \ref{fig:Ti44level}, by taking the average, the  nodal excited
state is plotted as a single state which we call the $0^+_2$ state. On this $0^+_2$ state the
nodal excited band is built. 

Around 12 MeV in excitation energy, another group of $1^-$ states having large $\alpha$
spectroscopic factors are reported \cite{Yam90,Yam93-2, Yam96}. Again, observed $\alpha$
spectroscopic factors are fragmented into several levels and the averaged value which we call the
$1^-_2$ state is shown in the figure. Although the assignment is not so firm,  another
negative-parity band is suggested on this state which we denote by ``negative-parity band
II''. The excitation energy of this band plausibly agrees with the cluster model calculation which
suggests the  $3\hbar\omega$ excitation 
of the intercluster motion. 

The information on the $\alpha$+$^{40}{\rm Ca }$ cluster states is summarized as
follows. First, the ground and the negative-parity band I built on the $1^-_1$ state at 6.2 MeV
constitute a parity doublet. Second, the nodal excited band built on the $0^+_2$ state around 11
MeV and the negative-parity band II on the $1^-_2$ state around 12 MeV may constitute another
parity doublet. The first doublet is dominated by the 0 and 1$\hbar\omega$ excitations of the
intercluster motion, while the second doublet is dominated by the $2\hbar\omega$ and $3\hbar\omega$
excitations.

\begin{table}[h]
\caption{The estimated intercluster distance of the ground and the $\alpha$+$^{40}{\rm Ca}$ cluster
states in units of fm, and the IS dipole and monopole transition matrix from the ground
state to the $1^-_1$, $1^-_2$ and $0^+_2$ states in units of $\rm fm^3$ and
$\rm  fm^2$. Intercluster Distances $D(0^{+}_2)$ and $D(1^{-}_2)$ are the averaged values of the
fragmented levels, while the transition matrix  $M^{IS0}(0^+_2)$ and $M^{IS1}(1^-_2)$ are sum of
them. Numbers in parenthesis are ratio to the Weisskopf  estimates. }\label{tab:Ti44ism} 
\begin{ruledtabular}
\begin{tabular}{ccccc}
 &$D_0$& $D_1$ & $D(0^+_2)$&$D(1^-_2)$   \\
\hline
 BB GCM &   5.5 & 6.0 & 7.0 & 7.5  \\
 AMD    &   2.5 & 5.0 & 6.0 & 7.0   \\
\hline
 &$M^{IS1}(1^-_1)$& $M^{IS0}(0^+_2)$ & $M^{IS1}(1^-_2)$&  \\
\hline
 BB GCM &   217.5 (11.7) & 47.2 (4.4) & 91.6 (4.9) & \\
 AMD    &   24.7 (1.3)   & 19.9 (1.8) & 16.7 (0.9) &  \\
\end{tabular}
\end{ruledtabular}
\end{table}

Next, we discuss the results of Brink-Bloch GCM. The Brink-Bloch GCM seriously overestimates
the energies of the observed cluster bands as well as the ground band indicating the increased
importance of  the cluster distortion. Because of too weak binding, all states  have large
intercluster distances that result in the huge dipole and monopole transition matrix listed in
Tab. \ref{tab:Ti44ism} which may be overamplified and unrealistic.

In the AMD results, the member states of the nodal excited band and negative-parity band II are
fragmented into several states as reported by the experiment. Therefore, for the states shown 
in Fig. \ref{fig:Ti44level} and the intercluster distances listed in Tab. \ref{tab:Ti44ism}, we
show the averaged values weighted by the amount of the cluster  component $S$ given by
Eq. (\ref{eq:overlap}). By taking the distortion effect into account, AMD gives a reasonable
description of the ground and cluster bands.  All states gain large binding energy compared to the
Brink-Bloch GCM, and  their intercluster distances, in particular that of the ground state, are
considerably reduced. This strong distortion is mainly due to the spin-orbit interaction and to
the formation of the mean field.  
\begin{figure}[h]
 \includegraphics[width=0.6\hsize]{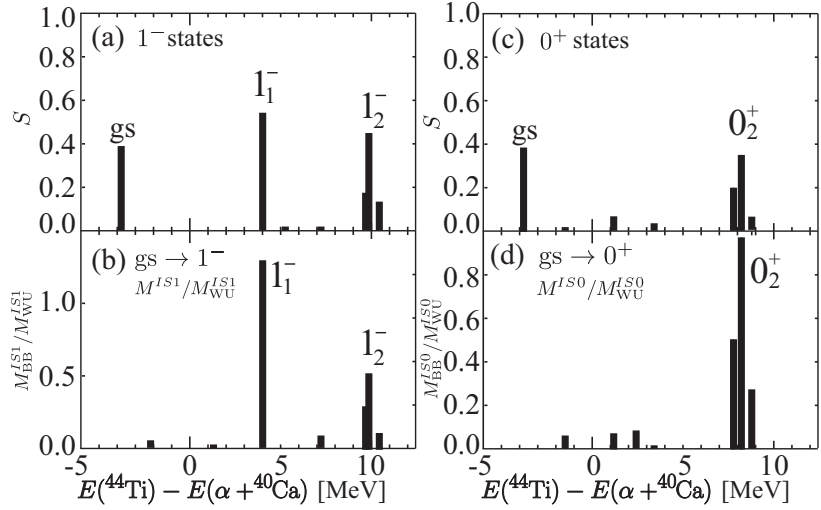}
 \caption{(a) The amount of the cluster component $S$ of the ground and $1^-$ states obtained by
 AMD. (b) The ratio of IS dipole transition matrix $M^{IS1}$ to the Weisskopf
 estimate. (c) Same as (a) but for the $0^+$ states. (d) Same as (b) but for the IS
 monopole transition.}\label{fig:Ti44isd}     
\end{figure}

Since the nodal excited band and negative-parity band II are fragmented, we discuss the transition
matrix by referring the distribution of the amount of the cluster component $S$. Fig
\ref{fig:Ti44isd} (a) shows $S$ of the ground and $1^-$ states as function of energy relative to
the $\alpha$ threshold. The ground  state has, despite of the strong cluster distortion,
considerable amount of cluster component $S=0.39$. However, we note 
that this does not necessarily mean the prominent clustering in the ground state. Most of this
cluster component is the wave function given in Eq. (\ref{eq:gswf}) and hence identical to the
shell model wave function. The first angular excited state (the $1^-_1$ state)  located at 4 MeV
above  the threshold has larger value of $S=0.59$. The second angular excited state (the $1^-_2$
state) is fragmented into three levels around 10 MeV above the threshold. If we sum up those
fragments, it amounts to $S=0.78$. The IS dipole transition matrix from the ground state to $1^-$
states are shown in Fig. \ref{fig:Ti44isd} (b). It is clear that both of the angular excited
states are strongly excited, because the IS dipole transition brings about 1 and 3$\hbar\omega$
excitation of the intercluster motion as shown by Eq. (\ref{eq:isd_decompose}). It must be noted
that many non-cluster $1^-$ states are obtained between the ground and $1^-_2$ states in AMD
calculation, but none of them have the transition matrix comparable with Weisskopf estimate. 
This result shows that the IS dipole transition has the sensitivity to 
for angular excited $1^-$ states, despite of the cluster distortion and fragmentation. 

As for the $0^+$ states and monopole transition, almost the same conclusion can be
drawn. Fig. \ref{fig:Ti44isd} (c) shows the amount of the cluster component in the $0^+$
states. Around 8 MeV, the nodal excited $0^+_2$ state is fragmented into three levels, which
amount to $S=0.6$. Again, we see the amount of cluster component $S$ and transition matrix are
strongly correlated. The nodal excited states have large IS monopole transition matrix, while
non-cluster states are insensitive. From those results, we 
conclude that the IS monopole and dipole transitions are good probe for asymmetric
clustering. Since both transitions can be measured simultaneously by the $\alpha$ inelastic
scattering, the experimental and theoretical survey look promising. 

We also comment on the giant resonances, which may exist at the similar energy region to the
$1^-_2$ and $0^+_2$ states. Because the peak of giant resonance may overlap with those states,
those nodal excited doublet may not be visible in the real situation. Nevertheless, we expect that
it is possible to identify those cluster state, because those cluster states will dominantly decay
by $\alpha$ emission, while the giant resonance decay by neutron emission.

\section{Summary}
In this study, we have discussed the IS dipole transition in $^{20}{\rm Ne}$ and
$^{44}{\rm Ti}$ that have $\alpha$+$^{16}{\rm O}$ and $\alpha$+$^{40}{\rm Ca}$ cluster states. 
In such asymmetric cluster systems, the existence of the angular excited $1^-$ cluster states is a
key to prove their asymmetric structure. We have shown that the isoscalar dipole
transition from the ground state strongly populates those asymmetric cluster states,
and hence, it is regarded as a good probe for such $1^-$ states. 

We first performed analytical calculations to estimate the magnitude of the transition
matrix. By rewriting the IS dipole operator in terms of the internal coordinates within clusters
and the intercluster coordinate, it was shown that the transition  brings about the 1 and
3$\hbar\omega$ excitation to the intercluster motion. Therefore, the IS dipole transition
has the potential to activate the degrees-of-freedom of cluster excitation embedded in the ground
state to populate the angular excited $1^-$ cluster states.

By assuming that the ground state is described by a shell model wave function, we have derived an
analytical expression of the IS transition matrix, and demonstrated that the transition
matrix is indeed enhanced and is in the order of Weisskopf estimate, even if the ground state has
an ideal shell model structure. We also performed a simple numerical calculation using Brink-Bloch
wave function to show that the transition matrix is amplified in order of magnitude if the ground
state has cluster correlation. 

To provide realistic and reliable result for IS monopole and dipole transitions  in
$^{20}{\rm Ne}$ and $^{44}{\rm Ti}$, nuclear structure calculations using Brink-Bloch GCM and 
AMD were performed. By taking the cluster distortion into account, AMD reasonably described the
energies of those cluster states. It was shown that despite of the cluster distortion, the nodal
and angular excited cluster states are strongly excited by the IS monopole and dipole
transitions, and hence, we conclude that the monopole and dipole transitions are promising probe
for asymmetric clustering.

\begin{acknowledgments}
Authors acknowledge the fruitful discussions with Dr. Kanada-En'yo, Dr. Kawabata and Dr. Bo. Zhou
were great help for this work. Part of the numerical calculations were performed on the HITACHI
SR16000 at KEK and YITP. This work was supported by the Grants-in-Aid  for Scientific Research on
Innovative Areas from MEXT (Grant No. 2404:24105008) and JSPS KAKENHI Grant Nos. 25400240 and
25800124.
\end{acknowledgments}
\appendix
\section{IS dipole operator represented by internal and intercluster coordinates} \label{app:appa}
We consider the $A$ nucleon system composed of the clusters with mass $C_1$ and $C_2$
($C_1+C_2 = A$), and wish to express $\mathcal M_\mu^{IS1}$ in terms of the internal
coordinates $\bm \xi_i$ within each cluster and the intercluster coordinate $\bm r$. Noting the
relations $\bm R_{C_1} - \bm r_{\rm cm} = {C_2}/{A}\bm r$ and $\bm R_{C_2} - \bm r_{\rm cm} =
-{C_1}/{A}\bm r$, the IS dipole operator is rewritten as follows,  

\begin{widetext}
\begin{align}
 \mathcal{M}_\mu^{IS1} =& \sum_{i=1}^{A}(\bm r_i - \bm r_{\rm cm})^2
 \mathcal{Y}_{1\mu}(\bm r_i - \bm r_{\rm cm})\nonumber\\
 =&\sum_{i\in C_1}\Bigl(\bm \xi_i + \frac{C_2}{A}\bm r\Bigr)^2
 \mathcal{Y}_{1\mu}\Bigl(\bm \xi_i + \frac{C_2}{A}\bm r\Bigr)
 +\sum_{i\in C_2}\Bigl(\bm \xi_i - \frac{C_1}{A}\bm r\Bigr)^2
 \mathcal{Y}_{1\mu}\Bigl(\bm \xi_i - \frac{C_1}{A}\bm r\Bigr) \nonumber\\
 =& \sum_{i\in C_1}\xi_i^2\mathcal Y_{1\mu}(\bm \xi_i)
 +\sum_{i\in C_2}\xi_i^2\mathcal Y_{1\mu}(\bm \xi_i)
 +\left(\frac{C_2}{A}\sum_{i \in C_1}\xi_i^2
 - \frac{C_1}{A}\sum_{i \in C_1}\xi_i^2\right)\mathcal Y_{1\mu}(\bm r)
 -\frac{C_1C_2(C_1-C_2)}{A^2}r^2\mathcal Y_{1\mu}(\bm r)\nonumber\\
 &+ 2\frac{C_2}{A}\sum_{i\in C_1}(\bm \xi_i\cdot\bm r) \mathcal Y_{1\mu}(\bm \xi_i)
 - 2\frac{C_1}{A}\sum_{i\in C_2}(\bm \xi_i\cdot\bm r) \mathcal Y_{1\mu}(\bm \xi_i),
 \label{eq:appa1}
\end{align}
\end{widetext}
where the relations 
$\sum_{i\in C_1} \bm \xi_i = \sum_{i\in C_2} \bm \xi_i = 0$ and 
$\mathcal Y_{1\mu}(\alpha\bm a + \beta\bm b)= \alpha\mathcal Y_{1\mu}(\bm a) + \beta\mathcal Y_{1\mu}(\bm b)$ 
are utilized. Using the identities, 
$\bm a\cdot \bm b=-4\pi/\sqrt{3}[\mathcal Y_{1}(\bm a)\otimes\mathcal Y_1(\bm b)]_{00}, 
 \ \ 
 [\mathcal Y_{1}(\bm a)\otimes\mathcal Y_{1}(\bm a)]_{1\mu}=0, \text{ and }
 [\mathcal Y_{1}(\bm a)\otimes\mathcal Y_{1}(\bm a)]_{2\mu}=
\sqrt{3/10\pi}\mathcal Y_{2\mu}(\bm a)$
 the term in last line of Eq. (\ref{eq:appa1}) reads,
\begin{align}
 (\bm \xi_i\cdot \bm r)\mathcal Y_{1\mu}(\bm \xi_i) =& -\frac{4\pi}{\sqrt{3}}
 [\mathcal Y_{1}(\bm \xi_i)\otimes[\mathcal Y_{1}(\bm \xi_i)\otimes\mathcal Y_1(\bm r)]_0]_{1\mu}\nonumber\\
 =&-\frac{4\pi}{\sqrt{3}}\sum_{l=0,1,2}\sqrt{2l+1}\left\{
 \begin{array}{ccc}
  1 & 1& l\\
  1 & 1& 0
 \end{array}
 \right\}
 [[\mathcal Y_{1}(\bm \xi_i)\otimes\mathcal Y_1(\bm \xi_i)]_l\otimes\mathcal Y_{1}(\bm r)]_{1\mu}
 \nonumber\\
 =&\frac{1}{3} \bm \xi_i^2\mathcal Y_{1\mu}(\bm r) - \sqrt{\frac{8\pi}{9}}
 [\mathcal Y_{2}(\bm \xi_i)\otimes \mathcal Y_1(\bm r)]_{1\mu}.\label{eq:appa2}
\end{align}
We see Eqs. (\ref{eq:appa1}) and  (\ref{eq:appa2}) yield
Eq. (\ref{eq:isd_decompose}).  

\section{Derivation of IS dipole matrix element}\label{app:appb}
Here, we derive Eq. (\ref{eq:redmat5}) from Eqs. (\ref{eq:redmat2}) and
(\ref{eq:isd_decompose}) in a similar way to Ref. \cite{Yam08}. First, we show that the first line
of Eq. (\ref{eq:isd_decompose}) identically vanishes in the case of the system composed of two
$LS$ closed shell (more strictly, $SU(3)$ scalar) clusters. This is easily proved by counting the
principal quantum numbers.  

For example, the first term of Eq. (\ref{eq:isd_decompose}) yields the matrix element proportional to  
\begin{align}
 \braket{\mathcal R_{N10}(\bm r)
 \Bigl(\sum_{i\in C_1}\xi_i^2\mathcal Y_{10}(\bm \xi_i)\phi_{1}\Bigr)\phi_{2}|\mathcal
 A'\Set{\mathcal R_{N_000}(\bm r)\phi_{1}\phi_{2}}}\nonumber.
\end{align}
Denoting the principal quantum number of $\phi_1, \phi_2$ as $N_{C_1}, N_{C_2}$,  the principal
quantum number of the ket state is equal to $N_0+N_{C_1}+N_{C_2}$. On the other hand, that of the
bra state is equal to or larger than $N+N_{C_1}+N_{C_2}+1$, because $\sum_{i\in
C_1}\xi_i^2\mathcal Y_{10}(\bm \xi_i)$ induces at least $1\hbar\omega$ excitation of
$\phi_{1}$. Since $N$ is equal to or larger than $N_0+1$, the principal quantum number of bra
state is larger than that of ket state, and hence, this matrix element vanishes. In the same way,
the third term of the first line yields 
\begin{align}
\langle\mathcal R_{N10}(\bm r)\mathcal Y_{1m}(\bm r)
 \Bigl(\sum_{i\in C_1}\mathcal Y_{2-m}(\bm \xi_i)\phi_{1}\Bigr)\phi_{2}
 |\mathcal
 A'\{\mathcal R_{N_000}(\bm r)\phi_{1}\phi_{2}\}\rangle.\label{eq:appb1}
\end{align}
The quantum number of $\sum_{i\in C_1}\mathcal Y_{2-m}(\bm \xi_i)\phi_{1}$ is at least
$N_{C_1}+2$, because  $\sum_{i\in C_1}\mathcal Y_{2-m}(\bm \xi_i)$ generates $2^+$ states of the
$LS$ closed shell nucleus $\phi_1$ which involves at least $2\hbar\omega$ excitation. Combined
with the quantum number of the intercluster motion which is at least $N-1$, we again find the quantum
number of the bra state is larger than that of the ket state. Thus, terms that involve the
internal cluster excitation vanish. 

However,  for the open-shell (non $SU(3)$ scalar) clusters, it must be noted that
Eq. (\ref{eq:appb1}) do not vanish and can be very large. A typical example is 
$^{12}${\rm C} cluster. For such clusters, the wave function in the parenthesis in
Eq. (\ref{eq:appb1}) is written as 
\begin{align}
 &\sum_{i\in {\rm ^{12}C}}\mathcal Y_{2-m}(\bm \xi_i)\phi_{\rm ^{12}C}(0^+_1)
 = \braket{\phi_{\rm ^{12}C}(2^+_1)| \sum_{i\in \rm ^{12}C}\mathcal Y_{2-m}(\bm \xi_i)|
 \phi_{\rm ^{12}C}(0^+_1)}\phi_{\rm ^{12}C}(2^+_1)
 +(\text{other excited $2^+$ states}). \label{eq:appb2}
\end{align}
Since $\phi_{\rm ^{12}C}(2^+_1)$ has the same principal quantum number with the ground state
$\phi_{\rm ^{12}C}(0^+_1)$ and the matrix element 
$\braket{\phi_{\rm ^{12}C}(2^+_1)| \sum_{i\in \rm ^{12}C}\mathcal Y_{2-m}(\bm \xi_i)|
 \phi_{\rm ^{12}C}(0^+_1)}$ is proportional to large $E2$ matrix element,
Eq. (\ref{eq:appb1}) can be comparable or even larger than the second line of
Eq. (\ref{eq:isd_decompose}). We conclude that if the cluster nucleus has the rotational or
vibrational ground band with enhanced $E2$ transition, the internal excitation of the cluster from
$0^+$ to $2^+$ can have large contribution to IS dipole excitation.

Now we evaluate the non-vanishing contribution from the second line. The first term of the second
line yields the the matrix element proportional to  
\begin{align}
 &\braket{\mathcal R_{N10}(\bm r)\mathcal Y_{10}(\bm r)
 \Bigl(\sum_{i\in C_1}\mathcal \bm \xi_i^2\phi_{1}\Bigr)\phi_{2}|\mathcal
 A'\Set{\mathcal R_{N_000}(\bm r)\phi_{1}\phi_{2}}}.\label{eq:redmat3}
\end{align}
Note that, in the bra state,  the IS monopole operator $\sum_{i\in C_1}\xi_i^2$ induces 0
or $2\hbar\omega$ excitation of $\phi_{C_1}$,  
\begin{align}
 \sum_{i\in C_1}\mathcal \bm \xi_i^2\phi_{1}
 =&\braket{\phi_1|\sum_{i\in C_1}\mathcal \bm \xi_i^2|\phi_1}\phi_1
 +(\text{$2\hbar\omega$ excited $0^+$ states} ). \label{eq:C1decomp}
\end{align}
and $\mathcal Y_{10}(\bm r)$ brings about the angular excitation of the intercluster motion
with $\pm 1\hbar\omega$, {\it i.e.,} the principal quantum number of the 
intercluster motion is equal to  $N\pm 1$. Again we count the quantum numbers and find that
Eq. (\ref{eq:redmat3}) is non zero only when $N=N_0+1$, otherwise the principal quantum number of
the bra state is larger than that of the ket state.    
From Eq. (\ref{eq:C1decomp}) and the following identities, 
\begin{align}
 &\mathcal R_{N10}(\bm r)\mathcal Y_{1m}(\bm r)
 = \sqrt{\frac{1}{4\pi}} r R_{N1}( r) Y_{00}(\hat r)
 +\sqrt{\frac{1}{5\pi}} rR_{N1}( r) Y_{20}(\hat r),\\
 &rR_{N1}(r) = \sum_{N'}\braket{R_{N'0}|r|R_{N1}}R_{N'0}(r),
\end{align}
Eq. (\ref{eq:redmat3}) is calculated as, 
\begin{align}
&\sqrt{\frac{1}{4\pi}}\braket{\phi_1|\sum_{i\in C_1}\mathcal \bm \xi_i^2|\phi_1}
 \sum_{N'}\braket{R_{N'0}|r|R_{N_0+11}}
 \braket{R_{N'0}(r)Y_{00}(\hat r)
 \phi_{1}\phi_{2}|\mathcal
 A'\Set{R_{N_00}(r)Y_{00}(\hat r)\phi_{1}\phi_{2}}}\nonumber\\
&=\sqrt{\frac{1}{4\pi}}C_1\braket{r_{C_1}^2}
 \braket{R_{N_00}|r|R_{N_0+11}}\mu_{N_0},
\end{align}
where $\braket{r^2}_{C_1}$ is the square of the root-mean-square radius of $\phi_{C_1}$.

Finally, the last term in the second line of Eq. (\ref{eq:isd_decompose}) yields
\begin{align}
 \braket{\mathcal R_{N10}(\bm r)r^2\mathcal Y_{10}(\bm r)
 \phi_1\phi_{2}|\mathcal
 A'\Set{\mathcal R_{N_000}(\bm r)\phi_{1}\phi_{2}}},\label{eq:redmat4}
\end{align}
where $r^2\mathcal Y_{10}(\bm r)$ brings about the nodal and angular excitations of the
intercluster motion with  $\pm 1\hbar\omega$ or $\pm 3\hbar\omega$, and hence the matrix element
vanishes except for $N=N_0+1$ and $N_0+3$ cases. By a similar calculation, one finds
Eq. (\ref{eq:redmat4}) is equal to 
\begin{align}
 \sqrt{\frac{1}{4\pi}} \braket{R_{N_00}|r^3|R_{N1}}\mu_{N_0},
\end{align}
where $N$ is  $N_0+1$ or $N_0+3$. 
From those results, we obtain an analytic expression for the reduced matrix element given in
Eq. (\ref{eq:isd_decompose}).

\section{Sign of $\bm f_{\bm N_0 \bf +1}$ and $f_{\bm N_0\bf +3}$}\label{app:appc}
The coefficients   $f_{N_0+1}$ and $f_{N_0+3}$ in Eq. (\ref{eq:ex1wf}) usually
have opposite sign for cluster states. To show it, we first approximate the wave
function of angular excited cluster state Eq. (\ref{eq:ex1wf}) as Brink-Bloch wave function given
in Eq. (\ref{eq:BB3}) and (\ref{eq:BB2}). This approximation may be justified, because those wave
functions have large  overlap to each other with proper choice of  $R$.
For example, in the case of $^{20}{\rm Ne}$, the overlap between the Brink-Bloch wave function
with $R=5.0$ fm and the AMD wave function for the $1^-_1$ state amounts to 82\%. 

Given that the approximation is reasonable, substitute Eq. (\ref{eq:BB4}) to Eq. (\ref{eq:BB2}) and
compare it with Eq. (\ref{eq:ex1wf}). As a result, one finds the sign of $f_N$ is equal to that of
$A_{Nl}$ defined by Eq. (\ref{eq:BB5}). Therefore, $f_{N_0+1}$ and $f_{N_0+3}$ should have
opposite sign, if the angular excited state is well approximated by Brink-Bloch wave function.

\end{document}